# Multiscale porosity in mesoporous bioglass 3D-printed scaffolds for bone regeneration

M. Natividad Gomez-Cerezo[a], Juan Peña[b,c], Sašo Ivanovski[a], Daniel Arcos[b,c], María Vallet-Regí[b,c] and Cedryck Vaquette[a]*

[a]The University of Queensland, School of Dentistry, Herston, QLD, Australia

[b]Departamento de Química en Ciencias Farmacéuticas, Facultad de Farmacia, Universidad Complutense de Madrid, Instituto de Investigación Sanitaria Hospital 12 de Octubre i+12, Plaza Ramón y Cajal s/n, 28040 Madrid, Spain

[c]CIBER de Bioingeniería, Biomateriales y Nanomedicina, CIBER-BBN, Madrid, Spain

## Abstract

In order to increase the bone forming ability of MBG-PCL composite scaffold, microporosity was created in the struts of 3D-printed MBG-PCL scaffolds for the manufacturing of a construct with a multiscale porosity consisting of meso- micro- and macropores. 3D-printing imparted macroporosity while the microporosity was created by porogen removal from the struts, and the MBG particles were responsible for the mesoporosity. The scaffolds were 3D-printed using a mixture of PCL, MBG and phosphate buffered saline (PBS) particles, subsequently leached out. Microporous-PCL (pPCL) as a negative control, microporous MBG-PCL (pMBG-PCL) and non-microporous-MBG-PCL (MBG-PCL) were investigated. Scanning electron microscopy, mercury intrusion porosimetry and micro-computed tomography demonstrated that the PBS removal resulted in the formation of micropores inside the struts with porosity of around 30% for both pPCL and pMBG-PCL, with both constructs displaying an overall porosity of 80-90%. In contrast, the MBG-PCL group had a microporosity of 6% and an overall porosity of 70%. Early mineralisation was found in the pMBG-PCL post-leaching out and this resulted in the formation a more homogeneous calcium phosphate layer when using a biomimetic mineralisation assay. Mechanical properties ranged from 5 to 25 MPa for microporous and non-microporous specimens, hence microporosity was the determining factor affecting compressive properties. MC3T3-E1 metabolic activity was increased in the pMBG-PCL along with an increased production of RUNX2. Therefore, the microporosity within a 3D-printed bioceramic composite construct may result in additional physical and biological benefits.

## 1. Introduction

Porous scaffolds for bone regeneration have been widely studied for many decades [1–3] due to the significant health burden of the great variety of conditions that require bone regeneration and reconstruction, ranging from traumatic fractures, resective surgery or other non-self-healing bony defects [4–6]. The general consensus is that scaffolds for bone tissue engineering must display an interconnected macroporosity, with pores sizes between 100 and 1000 micrometres [7,8], which facilitates angiogenesis and consequently bone ingrowth. In addition, the supply of nutrients and removal of metabolic waste through the macropores during *in vitro* maturation improves the success of those scaffolds after their implantation *in vivo* [7,9]. The presence of microporosity (considered here in the range of 10-100 microns) within the pores has been demonstrated to play a major role in the osteogenic capacity of the resulting scaffolds [10–14]. While the exact mechanism via which microporosity enhances bone formation is not fully understood, current knowledge implicates a physical effect resulting from higher capillary forces and an increased surface area available for cellular interaction and protein adsorption [15]. It has also been hypothesised that micropores may act as a reservoir for the accumulation of osteogenic biological cues which can trigger osteoblastic differentiation of progenitor cells [16].

Although, the effect of microporosity has been extensively studied in bioceramic scaffolds, principally fabricated from calcium phosphate biomaterials of different compositions [10–14], it has been less studied in 3D polymeric scaffolds. This may be because of the technical challenges encountered in developing 3D scaffolds with dual porosity. Although several studies have reported the fabrication of such scaffolds using a variety of scaffold fabrication techniques (ranging from negative mould imprinting, freeze-drying [17,18], or a combination of porogen leaching and freeze drying [19,20]), the reproducible and controlled manufacturing of a fully interconnected macro- micro-porous geometry remains challenging.

Additive manufacturing (AM), and more specifically 3D-printing, enables the reproducible fabrication of highly ordered scaffolds whereby the internal macroporous architecture is readily tailored and modifiable. 3D-printed constructs have been shown to be suitable for bone regeneration due to their excellent pore interconnectivity, which facilitates the revascularisation that is essential for osteogenesis [7]. However, the control over the surface microstructure and topography remains elusive, even though several reports have attempted to address this issue. One approach involved the utilisation of a solvent post-treatment, resulting in the partial dissolution of polymer on the strut surface, thus creating a micro-roughness that enhanced osteogenic differentiation [21,22]. Other methods have consisted of 3D-printing a viscous polymer solution on a cold collector plate prior to freeze-drying to create microporosity [23], or by printing in a coagulation bath favouring solvent non-solvent interaction for the formation of surface topography [24,25]. However, these post-treatment and manufacturing processes only creates microporosity in the submicron range, which is below the microporosity (ranging from 10 to 50 microns) that has been reported to favour osteogenesis.

In order to circumvent this limitation, Yun *et al* have proposed the combination of porogen leaching and 3D-printing, combining PCL and MBG. The technique involves the mixing of the porogen with the polymer, and extruding this mixture prior to leaching out the porogen in an aqueous solution. This approach resulted in the formation of micro-porosity within the struts of 3D-printed scaffolds in a controlled manner [26]. More recently, this technique has been further implemented to develop scaffolds with dual macro- and microporosity for drug delivery applications [27].

The highly ordered macroporous 3D-printed scaffolds manufactured using this technique possess intra-strut micropores whose dimensions are in the range of those beneficial for osteogenesis, as previously reported [10–14]. However, polymeric biomaterials are generally relatively inert, or at best osteoconductive, and the incorporation of bioceramic may be utilized as a strategy to enhance their osteogenic properties. In particular, mesoporous bioglass has been shown to stimulate osteoblast proliferation and differentiation, as well as promote cell spreading and colonisation [28–30]. This is due to the synergy between the bioglass chemical (composition and ions release) and textural properties (from their mesoporous nature), which enhances their bioactivity both *in vitro* and *in vivo* [28,31,32]. Therefore, the manufacturing of a bioactive scaffold, possessing nanoscale topographical features capable of facilitating initial cell differentiation, microporosity required for the accumulation of ions that promote biomineralisation, and a macro-porous network facilitating the establishment of vascularization, represents a sound strategy for bone regeneration.

Notably, although additively manufactured scaffolds featuring dual porosity have been previously developed as a proof of concept [26], little is known about their porosity formation, bioactive properties, and biological behavior.

In order to address these knowledge gaps, we prepared multiscale meso- micro- macro-porous MBG composite scaffolds, and studied comprehensively their physical properties, the effect of porogen leaching as well as the effect in the construct and the strut porosity and mechanical properties. In addition, their biomineralisation capacity and osteogenic potential in vitro.

## 2. Materials and methods

### 2.1. Synthesis of mesoporous bioactive glass MBG-58S

In this study the use of mesoporous bioglass was preferred over bioglass due to the enhanced bioactivity that the mesoporosity imparts to this grade of biomaterials as previously demonstrated [33].

Mesoporous bioactive glass MBG-58S with composition 58 $SiO_2$– 37 CaO – 5 $P_2O_5$ (% mol) was synthesised by evaporation induced self-assembly (EISA) [34,35], using the nonionic surfactant Pluronic F127 (polyoxyethylene–polyoxypropylene–polyoxyethylene type polymer $(PEO)_{100}$-$(PPO)_{60}$-$(PEO)_{100}$) as a structure-directing agent. F127 (2 g) were dissolved in 38 g of ethanol with 1 mL of 0.5 M HCl solution. Subsequently, 3.75 mL of tetraethyl orthosilicate (TEOS), 0.44 mL of triethyl phosphate (TEP) and 2.47 g of $Ca(NO_3)_2.4H_2O$ were added under continuous stirring at 3 h intervals. The obtained sols were cast in Petri dishes (9 cm diameter) to undergo the EISA process at 30 °C for 7 days. The resulting gels were treated at 700 °C for 3 h to obtain the final MBG-58S as a calcined glass powder. The powder was sieved and the particle size below 40 µm was collected. All reactants were purchased from Sigma-Aldrich and used without further purification.

### 2.2. Preparation of mesoporous bioactive glass/polycaprolactone scaffolds

The fabrication of the microporous 3D-printed scaffolds involved a two staged protocol as previously reported [36]. The first step consisted of mixing the porogen and/or the MBG with a polycaprolactone (PCL, CAPA6400, Mw=37,000 Da, TRIISO) solution. To this end, 100mL

PCL solution (10% wt/vol) was prepared and 40 microns sieved PBS particles (obtained from PBS solid tablets, ThermoFisher) were added as porogen with or without MBG particles.

Three different types of scaffolds were fabricated by gradually adding 30% wt/wt (PBS) to the PCL solution (porous PCL, **pPCL**), PBS (30%) and MBG-58S (20%) (**pMBG-PCL**) or only MBG-58S (20%) (**MBG-PCL**, non-microporous). Then, the solutions were stirred for 3 hours and sonicated for 1 hour until a homogeneous suspension was obtained, the mixture was cast in a glass dish to allow for chloroform evaporation at room temperature. The PCL-MBG-PBS ratio was selected based from a previous report and based on the extrusion capability and reproducibility of the resulting mixture in our 3D-printer [36].

The three different groups were 3D-printed using an in-house bioextruder under the following conditions: the PCL-PBS, PCL-PBS-MBG or the PCL-MBG blends were placed in a stainless steel reservoir, heated at 140°C through a heated cartridge unit and extruded through a metal nozzle (inner diameter of 0.58 mm) on a programmable stage translating at 280 mm/min, using a 0/0/90/90 pattern with a layer height of 0.3 mm and a strut interdistance of 2mm. The scaffolds were subsequently soaked in NaOH (0.01M) at 37°C for 2 weeks to leach out the porogen with the solution replaced every 3 days; this treatment removed the PBS and consequently produced microporosity in the struts. Then, the scaffolds were rinsed 3 times with deionised water and then kept in water for 12 hours at 37ºC.

**2.3 Thermogravimetric analysis**

In order to determine the relative amount of PCL, MBG and remaining PBS, thermogravimetric analysis (TGA) was carried out using a TG/DTA Seiko SSC/5200 thermobalance (Seiko Instruments, Chiva, Japan) between 30°C and 650 °C in an air atmosphere at a heating rate of 1°C.min$^{-1}$, using platinum crucibles and α-Al2O3 as references.

**2.4 Porosity –N2 adsorption**

The textural properties MBG particles were determined by nitrogen adsorption with a Micromeritics ASAP 2020 equipment (Micromeritics Co., Norcross, USA). Prior to the N2 adsorption measurements, the samples were previously degassed under vacuum during 24 h, at 105 °C. The surface area was determined using the Brunauer-Emmett-Teller (BET) method. The pore size distribution between 0.5 and 60 nm was determined from the adsorption branch of the isotherm by means of the Kruk-Jaroniec-Sayari Standard (KJS) method.

**2.5 Transmission electron microscopy (TEM)**

TEM was carried out using a JEOL-3010 microscope, operating at 300 kV (Cs 0.6 mm, resolution 1.7 Å). Images were recorded using a CCD camera, using a low-dose condition.

**2.6 Scanning electron microscopy (SEM)**

Scanning electron microscopy (SEM) was carried out using a JSM F-7001 microscope (JEOL Ltd., Tokyo, Japan). The scaffolds (n=2) from each immersion time point were mounted onto SEM stubs and carbon coated in vacuum using a sputter coater (Balzers SCD 004, Wiesbaden-Nordenstadt, Germany). Energy-dispersive X-ray spectroscopy (EDS) measurements on each of the two scaffolds were taken for each group (n=2 with 6 measurements in each sample).

**2.7 Porosity – Microcomputed Tomography (Micro-CT)**

Both the macro and micro porosity of the scaffolds were assessed using Micro-CT. The samples were scanned using a mCT40 (SCANCO Medical AG, Brüttisellen, Switzerland), applying the following conditions: 55 kVp, 145 µA, 8 W, a voxel size of 30 µm, at a grayscale threshold of 150. Quantification of the overall porosity was performed by segmenting the entire scaffold while the microporosity was measured by segmenting an area of interest within the struts (n=5 in both measurements). Three-dimensional images of the scaffolds were reconstructed by the software package.

**2.8 Porosity – Mercury Intrusion Porosimetry**

Scaffolds were studied by Mercury Intrusion Porosimetry in order to measure the strut porosity, pore diameter and surface area. This analysis was done with a Micromeritics Autopore IV 9500 device (Norcross, GA, USA).

**2.9 Ion release study**

The scaffolds 8 x 8 x 3 mm$^3$ (n=3) were immersed for 15 days in 10 mL of 0.01 M NaOH at 37°C and the entire volume of NaOH was collected every 3 days in order to determine the ion release using Atomic Emission Spectroscopy inductively coupled plasma (ICP-OES). The analysis of Ca and P released was quantified through the emission lines 317.93, 251.61 nm respectively, on a Varian model view AX Pro (Varian Inc, Palo Alto, CA, USA). The detected concentrations were in the calibration range between 0.1 and 10 mmol/L. The measurements were performed in the corresponding emission range of equipment (167-758 nm) as previously reported [30,37].

## 2.10 Fourier transform infrared spectroscopy (FTIR)

Fourier transform infrared (FTIR) spectroscopy was carried out in a Nicolet Nexus (Thermo Fisher Scientific) equipped with a Goldengate attenuated total reflectance device (Thermo Electron Scientific Instruments LLC, Madison, WI USA) (n=3).

## 2.11 Nano-indentation

The surface hardness was measured using a Hysitron TI 950 Triboindentor (BRUKER) with a conical tip 60º, 5 µm (two different scaffolds of dimension 8 x 8 x 3 mm$^3$ per group post-immersion in NaOH for 15 days were utilised with 3 measures per scaffold, n=2 in triplicate). In order to obtain accurate measurements, the analysis was performed on the scaffold area facing upwards, where the struts are flatter, as can be observed on the SEM images. The scaffolds were vacuum dried overnight and glued into the holder prior to the analysis.

## 2.12 Compressive Young's modulus

Mechanical compression tests were performed using an Instron 5848 microtester with a 500 N load cell (Instron Australia) in PBS at 37 °C. Scaffolds (n=6), with dimensions of 8 x 8 x 3 mm$^3$, were subjected to 50% of compression at a rate of 1mm/min. The compressive Young's modulus of the scaffolds was calculated using the initial linear portion of the stress versus strain data.

## 2.13 Assessment of *in vitro* bioactivity

Individual scaffolds (50 mg) (n=3) of each group were soaked in 10 mL (%wt/vol = 0.5%) of filtered simulated body fluid (SBF pH=7.4) [38] based on our previously published protocol [30], in polyethylene containers at 37° C under sterile conditions for 3 days. SEM and FTIR were used to study the evolution of the surface of the composite scaffolds.

## 2.14 Cell culture

The purpose of the *in vitro* cell culture study was to evaluate the effect of the microporosity and the presence of the MBG particles on osteogenicity. The same three scaffold groups were evaluated: 1) pPCL, 2) pMBG-PCL, and 3) MBG-PCL. Mice pre-osteoblastic cells (MC3T3-E1) were cultured in T75 flasks using DMEM medium supplemented with 10% (v/v) of fetal bovine serum (FBS) and 1% (v/v) penicillin/streptomycin (basal medium). Prior to cell seeding, all scaffolds were sterilised by immersion in 80% ethanol for 15 min and UV-sterilization for 20 min on each surface. Thereafter, the scaffolds were immersed in 1mL of basal medium overnight to allow for protein adsorption.

50,000 cells at passage 5 in 20 µL of medium were seeded in the scaffolds and cell adhesion was allowed for 6 hrs. A regular hydration step was performed every 30 min for the first 2 hours and then every hour for another 2 hrs. Six hours post seeding, 1 mL of medium was added to the well and the culture took place over 7 days with a medium change every 2 days.

## 2.15 Cell metabolic activity

At days 4, 7 and 14, the cell metabolic activity on the various scaffolds (n=5 for each group and each time point) was quantified using Alamarblue®. The scaffolds were transferred to new wells and 400 µL of 10% Alamarblue® solution were added. The scaffolds were further incubated for 4 hrs at 37°C under $CO_2$ (5%) atmosphere and then exposed to the resazurin solution. Thereafter, the Alamarblue® solution was removed, the scaffolds were rinsed in PBS, fresh medium was added to the wells and the culture was continued until a given time point (4, 7 or 14 days). Then, the fluorescence signal was read from 150 µL of solution in triplicate at $\lambda_{em} = 590$ nm using a $\lambda_{exc} = 560$ nm with a fluorescence spectrometer Biotek Synergy 4. The percentage of reduction of the resazurin, indicating the level of cell metabolism, was calculated according to the manufacturer's instructions.

## 2.16 ALP activity

The alkaline phosphatase (ALP) activity of cells cultured onto the scaffolds after 7 and 14 days was assessed as a marker of early osteogenic differentiation. For this experiment, osteogenic medium was prepared by supplementing with β-glycerolphosphate (50 mg/mL, Sigma Chemical Company, St. Louis, MO, USA) and L-ascorbic acid (5 mM, Sigma Chemical Company, St. Louis, MO, USA). After 7 and 14 days, the scaffolds (n= 4 for each group) were lysed in 500 µL of a 0.1% Triton X-100 solution for 1 hr. Thereafter, 50 µL of the ALP lysate was added in 750 µL of p-nitrophenylphosphate solution (10 mmol/L) and incubated for 30 mins at 37°C. The reaction was subsequently stopped by the addition of 50 µL of 1 M NaOH. The absorbance at 410 nm of 100 µL of this solution was measured in triplicate using a Helios Zeta UV–vis spectrophotometer and normalised by the total protein content (measured by BCA kit from Sigma Aldrich at 560 nm).

**2.17 Cell morphology**

Cells were cultured in basal medium for 7 and 14 days on the scaffolds, which were rinsed twice in PBS and fixed with 4 % glutaraldehyde in PBS for 6 hours. After rinsing 3 times with PBS, dehydration was carried out by rinsing 3 times with graded ethanol solutions (30, 50, 70, 90 and 95%), with a final dehydration in absolute ethanol. The samples were vacuum-dried overnight, mounted on stubs and gold coated using a sputter coater (Balzers SCD 004, Wiesbaden-Nordenstadt, Germany). Scanning electron microscopy was then carried out with a JSM F-7001 microscope (JEOL Ltd., Tokyo, Japan), operating at 5 kV.

**2.18 Morphological studies by confocal laser scanning microscopy**

Confocal microscopy was performed to visualise the morphology of the cells attached onto the scaffolds cultured in basal medium after 7 and 14 days. Each scaffold was rinsed twice in PBS (Polysciences, Warrington, PA, USA) and fixed in 4% (w/v) paraformaldehyde in PBS with 1% (w/v) sucrose at 37°C for 4 hrs. Thereafter, the scaffolds were rinsed twice with PBS and then the samples were incubated 5 min with Triton 0.1 % (w/v) at room temperature to permeabilise the cells. Then, the scaffolds were rinsed with PBS and incubated with 400 µL, 6-diamidino-2-phenylindole (DAPI) (DAPI, Vector Laboratories, Burlingame, CA, USA) (1:1000) and Atto 565-conjugated phalloidin at a concentration of 0.165 µM (Molecular Probes) for 30 min. DAPI (cell nuclei) and phalloidin (cytoskeleton) were visualised in blue and red, respectively.

**2.19 Immunostaining for the osteogenic marker RUNX2**

The expression of Runt-related transcription factor 2 (RUNX2) by the osteoblasts seeded on the 3D printed scaffolds was detected using immunostaining. The osteoblast seeded scaffolds were cultured for 14 days and fixed with 4% (w/v) paraformaldehyde for 30 min. The scaffolds were then washed with PBS and permeabilised using a solution containing 0.2% (v/v) Triton X-100, 320 mM sucrose and 6 mM $MgCl_2$ in PBS for 30 min. Thereafter, the scaffolds were treated with blocking solution containing 1% (w/v) bovine serum albumin and 0.3 M glycine in PBS for 1 h and incubated with 1 ng/µl of primary antibody (mouse anti-RUNX2 from Santa Cruz Biotechnology, Inc, USA) overnight at 4 °C. After PBS washing, the scaffolds were then stained with a solution containing 5 ng/µl Alexa Fluor® 488-conjugated-anti mouse antibody (Abcam, UK), DAPI (5 µg/ml) and Atto 565-conjugated phalloidin (0.8 U/ml) for 1 hour in PBS at room temperature protected from light. The stained samples were washed with PBS and imaged using a Nikon ECLIPSE Ti confocal microscope. The relative quantification of RUNX2 was carried out calculating the area in green, corresponding to RUNX2 normalised to the area in blue, which represents the total number of cells.

**2.20 Statistical analysis**

Data is expressed as means ± standard deviations, and the statistical analysis was performed using the Statistical Package for the Social Sciences (SPSS) Software using one-way ANOVA followed by a Tukey post hoc test for multiple comparisons.

## 3. Results and discussion

**Figure 1** and **table 1** summarised the morphologic and the textural characterisation of MBG-58S. The porosity was studied by $N_2$ adsorption and TEM (Figure 1A, 1B and 1C), showing a highly porous structure at the nanoscale. The MBG particles were ground and sieved, and their morphology and size was observed by SEM (Figure 1D), confirming a size was under 40 micrometres. Once blended with the PCL, the mixture was 3D-printed using a superimposed strut design with a 0/90 layer by layer pattern (figure 1E and 1F), thus resulting in an arrangement of highly interconnected tetragonal pores with sizes around 1 mm in the x,y plane and 500 micrometers along z axe.

**Table 1.** Textural parameters obtained by N2 adsorption for MBG-58S.

| BET Surface ($m^2/g$) | Pore Volumen ($cm^3/g$) | Pore Diameter (nm) |
|---|---|---|
| 149.1 | 0.14 | 5.6 |

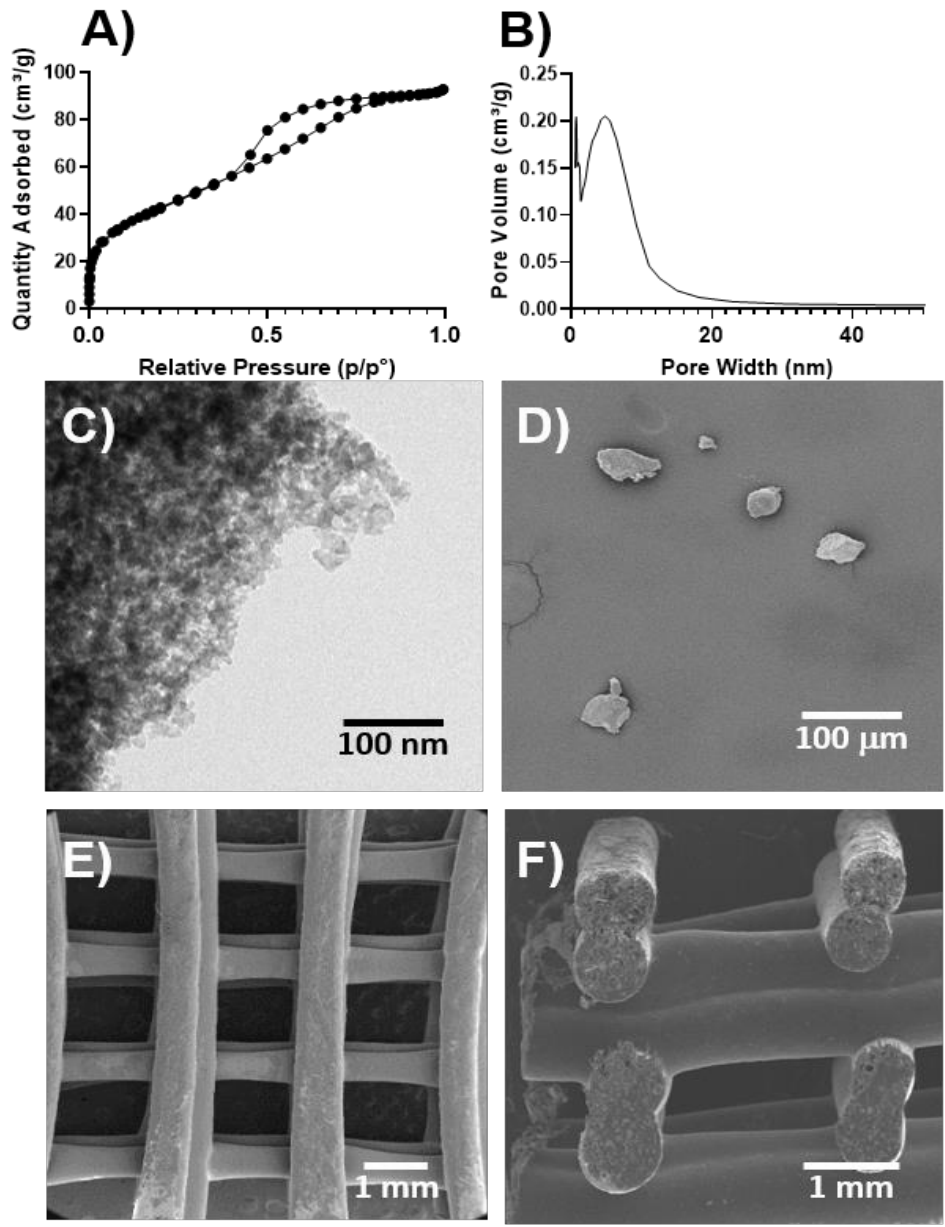

**Figure 1.** A) Nitrogen adsorption/desorption isotherms of MBG-58S B) pore size distribution of MBG-58S. Microscopy characterization of C) TEM of MBG-58S, D) SEM of MBG-58S in particles and SEM of the 3D-printed construct E) bottom view,

**Figure 2** shows the surface differences of the three kinds of scaffolds before and after leaching out. The pPCL scaffolds before leaching displayed a smooth surface typical of 3D-printed PCL scaffolds and some PBS particles were observed on the surface. The cross-section of the pPCL scaffolds pre-leaching was not micro-porous, although occasional large spherical pores were observed, which were created by the entrapment of air bubbles in the PCL-PBS slurry during the extrusion (Figure 2 as indicated by the *). The PBS particles were also observed in the bulk of the strut (Figure 2 as indicated by the white arrows). Similar morphology was observed for the other groups (MBG-PCL and pMBG-PCL) prior to immersion in 0.01 M NaOH, displaying here again large spherical pores corresponding to the entrapment of the air bubbles which is consistent with a previous report using this technique [36]. Upon PBS removal, the surface morphology of the microporous scaffolds changed significantly, revealing the presence of pores on the surface of the pPCL scaffolds as well as within the bulk of the struts, as shown in the cross-sectional images. Similarly, the pMBG-PCL scaffolds also displayed pores on the surface of the struts. The presence of pores on the surface of the micro-porous groups (pPCL and pMBG-PCL) is an important feature for enabling cell and tissue colonisation within the bulk of the strut within the microporous network formed by the removal of the PBS. Therefore, the resulting scaffolds possessed a hierarchical porous structure, including a microporous network which can potentially improve cell infiltration and osteogenesis [39] and a highly ordered macroporous network which enables blood clot formation and stabilisation, favourable for neo-vascularisation, a pre-requisite for osteogenesis [7,40].

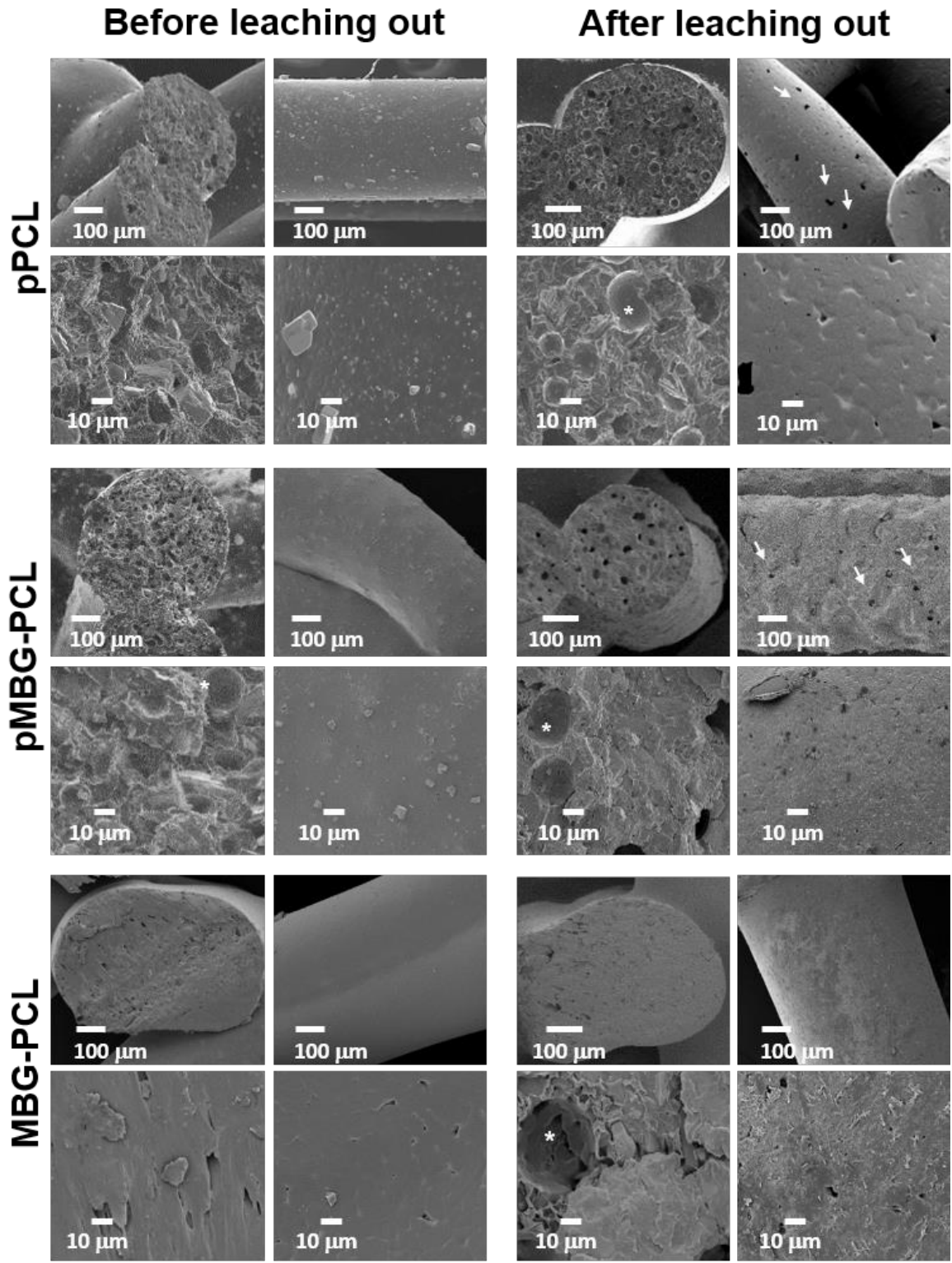

**Figure 2.** SEM images before and after the salt-leaching out in NaOH. White arrows show the porosity created after the treatment and stars show spherical air bubbles formed during the printing.

Interestingly, the pMBG-PCL scaffolds underwent biomineralisation in the form of a calcium phosphate layer deposition during the PBS removal step as demonstrated by the rougher surface of the struts and the presence of inorganic material on the pore's walls. This is also of interest for the enhancement of the osteogenic properties of the scaffolds, as surface topography and composition are important factors contributing to osteoblastic differentiation [21,41–43]. However, the MBG-PCL scaffolds did not display any biomineralisation during PBS removal. This was further demonstrated by EDXS, which revealed that the pPCL and MBG-PCL had no detectable phosphorus or calcium on their surface whereas these ions were measured with a Ca/P ratio of 1.79 on the pMBG-PCL. The spontaneous biomineralisation of the MBG 3D-printed scaffold, when immersed in PBS, was previously reported by our group and is a potent means of modifying the surface features and bioactivity of MBG/polymer construct [44]. Our previous report demonstrated that the immersion of MBG particles in a medium rich in phosphate ions, such as PBS, resulted in modification of the topography, composition and mechanical properties of 3D-printed scaffolds, and ultimately was beneficial for in vitro osteoblastic differentiation. In addition, mineralisation was found inside the micropores of the pMBG-PCL constructs, showing a Ca/P ratio about 1.79, similar to native hydroxyapatite (HA) that present a Ca/P ratio about 1.67 (**Figure 3**) while bare polymer surface was only visible in the other groups.

In order to verify the removal of the PBS particles, the scaffold mass loss was investigated throughout the leaching out step (as shown in **table 2**). A mass loss of about 30 % was demonstrated in the pPCL scaffolds, hence matching the theoretical PBS amount introduced in the manufacturing process. Similar results were observed in the pMBG-PCL scaffolds, whereas no measurable mass loss was detected in the case of the non-microporous MBG-PCL scaffolds.

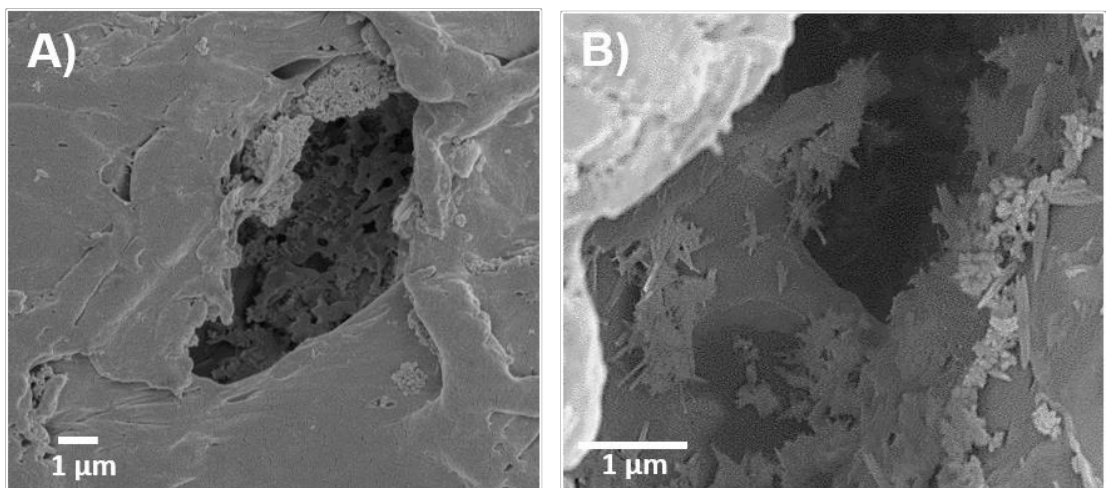

**Figure 3.** pMBG-PCL SEM image after PBS-leaching showing the pore structure and the early mineralisation deposited on the proe's wall. A) x5000 and B)x10000.

The removal of the PBS was further confirmed using TGA, which demonstrated that less than 1% wt/wt of the porogen remaining post-leaching, in accordance with previous reports [27,36]. The remaining quantity of ceramic in the pMBG-PCL scaffolds was approximately 23%, whereas 20% inorganic materials remained in the MBG-PCL scaffolds. This indicates that the formation of the microporosity created a suitable microarchitecture for the deposition of the mineralised layer, which accounts for the increased bioceramic content.

**Table 2.** Mass loss and TG analysis after the leaching out.

| Sample After 15 days in NaOH (0.01M) | Mass loss (%) | TG mass loss (%) | MBG quantity (%) |
| --- | --- | --- | --- |
| pPCL | 29.8 ± 3.7 | 99.01 | - |
| MBG-PCL | No mass loss | 80.34 | 19.66 |
| pMBG-PCL | 30.5 ± 1.5 | 77.73 | 22.27 |

The porosity of the scaffolds or of individual struts was investigated by microcomputed tomography. Via its differential segmentation (that is, either the whole scaffold or individual struts), this technique enabled the quantification of the overall macroporosity of the constructs and the microporosity within the struts. The 3D reconstruction of the pPCL and pMBG-PCL scaffolds before and after leaching further confirmed the apparent increased roughness on the surface of the scaffolds and struts, as shown in **Figure 4A**, whereas the MBG-PCL remained mostly unchanged. The quantification of the overall porosity of the various constructs demonstrated a 10% increase in the porosity of the microporous scaffolds (PBS-leached) regardless of the presence of the MBG particles. Here again the porosity of the non-microporous MBG-PCL did not change significantly and remained around 70% (**Figure 4B**). This was consistent with a 10% increase in the porosity of a MBG-NaCl leached 3D-printed scaffold as previously reported [26] albeit for a slightly higher porogen ratio. As previously reported the ratio porogen/ polymer will determine the resulting porosity of the scaffold [45] and struts. However in the context of additively manufactured microporous scaffolds, the amount of porogen that can be incorporated is limited by the rheological behaviour of the mixture that must allow proper extrusion in the printing head along with the scaffold physical integrity post-porogen removal.

When considering the intra-strut microporosity, all groups prior to leaching displayed a porosity in the range of 5%, which is most likely the contribution of the entrapped air bubbles in the molten mixture during extrusion. As expected, the removal of the PBS significantly increased the intra-strut porosity to around 30% for the pPCL scaffolds which is consistent with a previous report [27] and 20 % for the pMBG-PCL (**Figure 4C**). The deposition of the biomineralised layer within the pores of the pMBG-PCL scaffolds may account for this discrepancy as the layer will occupy some volume, consequently reducing the porosity.

The intra-strut porosity of the various groups was further assessed using Hg intrusion porosimetry, as shown in **table 4**. This confirmed that the microporous scaffolds displayed an increased intra-strut porosity ranging from 20 to 30% for both pPCL and pMBG-PCL groups, whereas it remained relatively low (6%) for the MBG-PCL, which was in the range of the µCT measurements. This further demonstrated that pore dimensions from 0.05-50 µm, corresponding to the PBS particle sizes, were the major contributor of the intra-strut porosity around 20% for pPCL and 25% for pMBG-PCL. This extra 5% of porosity in pMBG-PCL is probably related to the space between the glass particles, corresponding with the percentage of this porosity in the MBG-PCL. The large pores associated with the presence of air entrapped in the polymer blend mixture only contributed to a few percent of the global porosity. As a validation of the experiment the total volume of intruded mercury is presented in **table 3**, indicating that a higher amount of mercury was intruded in the micro-porous specimens. This fluid penetration showed that any aqueous medium could migrate within the struts due to the microporous network.

**Table 3.** Porosity and intruded Hg for the scaffolds after the leaching out process.

|  | POROSITY (%) | | | INTRUDED MERCURY (mL/g) |
|---|---|---|---|---|
|  | TOTAL 0.05-400 µm | 0.05-50 µm | 50-400 µm | TOTAL 0.05-400 µm |
| pPCL | 23.3 ± 2.1 | 19.8 ± 1.78 | 3.5 ± 0.3 | 0.30 ± 0.01 |
| pMBG-PCL | 30.0 ± 2.7 | 25.4 ± 2.2 | 4.6 ± 0.4 | 0.34 ± 0.02 |
| MBG-PCL | 6.5 ± 0.6 | 4.3 ± 0.3 | 2.2 ± 0.2 | 0.05 ± 0.002 |

FTIR was performed to study the biomineralisation that spontaneously occurred on the surface of the scaffolds and in the walls of the micropores during the leaching out process (**Figure 5A**). The scaffolds were studied before and after the leaching out in NaOH and the pPCL and MBG-PCL scaffolds showed similar spectra pre- and post-leaching, indicating that no changes had occurred in these constructs during the leaching out step. However, in the pMBG-PCL scaffolds, the presence of a doublet band at 600 cm$^{-1}$ demonstrated that the biomineralised layer observed on the surface was composed of crystalline calcium phosphate [46–48], which was consistent with our previous report [44].

The formation of this calcium phosphate layer can be attributed to the simultaneous release of phosphorus, principally from the PBS, and calcium from the MBG particles. Therefore, the release profile of these ions was investigated using ICP-OES. The various groups displayed significantly different release profiles. Indeed, the pPCL shows a phosphorus release of about 60 ppm (**Figure 5 B**), corresponding to the quantity of P in the PBS added in the scaffolds. As expected, no release of calcium was detected (**Figure 5C**), since no calcium was present in the initial pPCL scaffold composition. The MBG-PCL scaffolds showed a very low release of phosphorus due to the low concentration of phosphate in the MBG particles. The MBG-PCL scaffolds exhibited the highest level of calcium release when compared to the other groups. However, the calcium release remained very low (below 4 ppm) throughout the 15 days of leaching out, which is markedly different from other studies [49–51]. This is most likely caused by the slight basic conditions into which the construct was immersed, as opposed to buffered or simulated body fluid media [52–54]. In addition, the ion release from MBGs and PCL-MBGs scaffolds in DMEM or α-MEM was also studied. The release pattern was slower than in SBF, highlighting the essential role that the composition and pH play in ions exchange with the media [30,55–57].

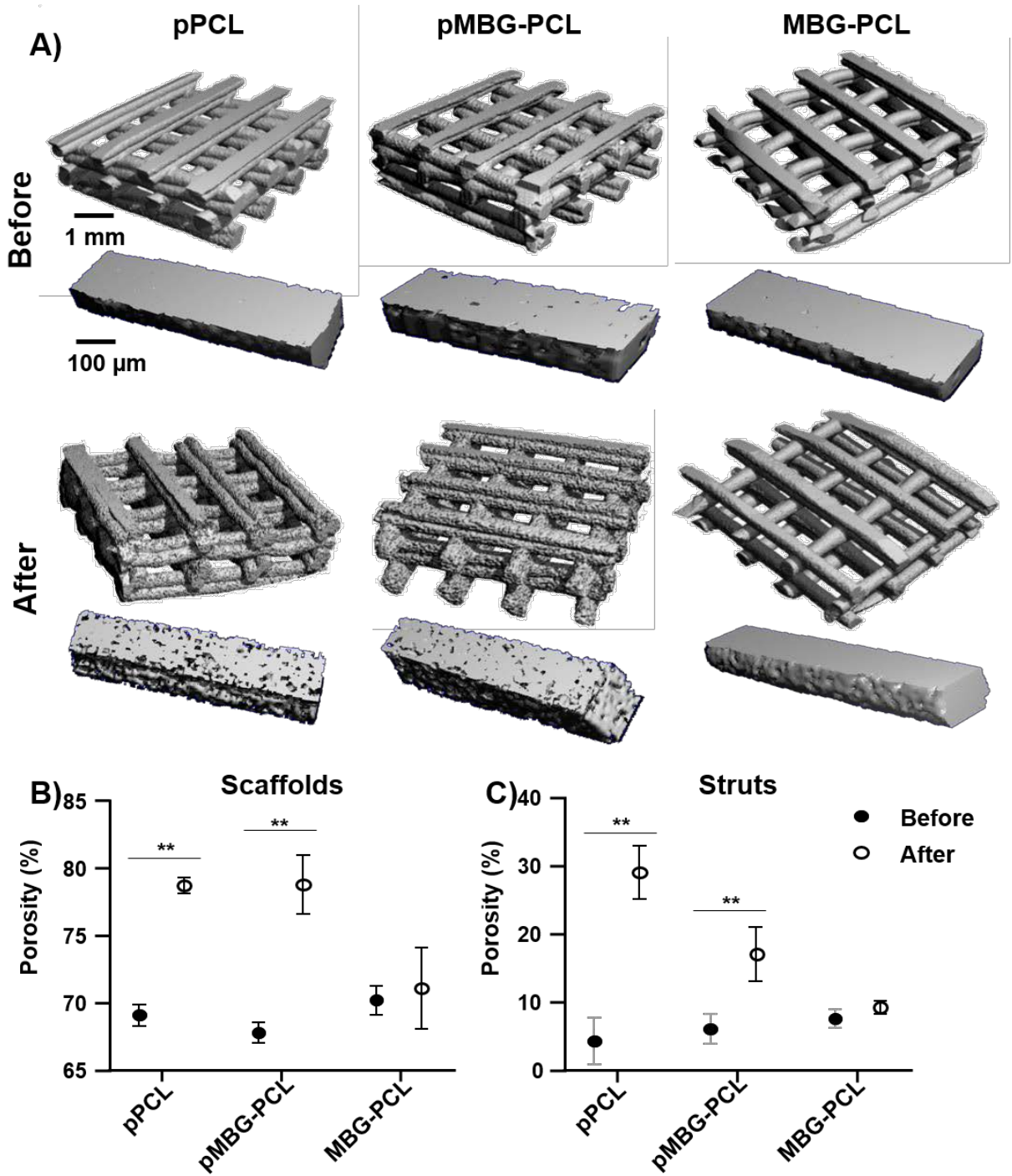

**Figure 4.** Microcomputed tomography (μCT), A) reconstruction of scaffolds and struts before and after the leaching out process, B) scaffolds porosity, C) struts porosity. The stars represent statistical significance (p < 0.01) (n=5).

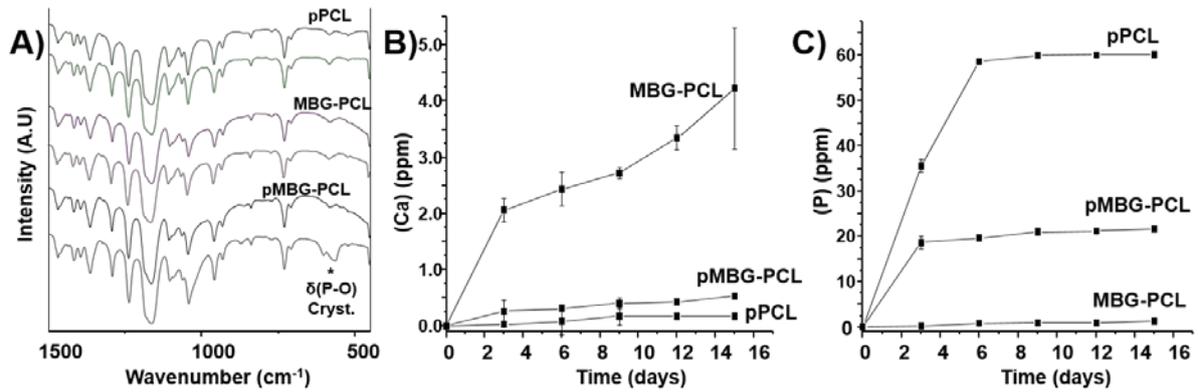

**Figure 5**. A) FTIR before and after the leaching out, B) Ca release during the PBS leaching out, C) P release during the PBS leaching out.

The pMBG-PCL scaffolds showed a constant release of P (20 ppm) and Ca (0.2 ppm) after 3 days, with concentrations in the mid-range of those observed for the pPCL and MBG-PCL scaffolds for the corresponding ions. The very low level of calcium released from the pMBG-PCL, combined with the differential amount of P released when compared to the pPCL, indicates that these ions were consumed immediately after release to form the biomineralised layer previously observed. A similar trend in the release profile of these ions was observed in our previous report, albeit using a water-soluble polymer in the 3D-printed composite [44].

The impact of biomineralisation and PBS removal on biomechanical properties was investigated using nano-indentation to measure the surface hardness, and at a macroscopic level by compression testing (**Figure 6**). The formation of the biomineralised layer on the pMBG-PCL scaffolds resulted in a 3- and 2-fold increase in surface hardness when compared to the pPCL and MBG-PCL scaffolds respectively (**Figure 6A**), which is consistent with our previous report [44]. In terms of macroscopic properties, the formation of microporosity within the scaffold significantly reduced the compressive strength of the 3D-printed scaffolds (**Figure 6B**). The pPCL, MBG-PCL and pMBG-PCL scaffolds had compressive moduli of 6, 5 and 28 MPa respectively, which was in accordance with what was previously reported for microporous [26,36] and non-microporous scaffolds [30,58]. As a result of the incorporation of microporosity, the compressive modulus of both pPCL and pMBG-PCL was significantly reduced when compared to the MBG-PCL scaffolds (**Figure 6B**) or compared to non-microporous 3D-printed PCL scaffolds reported in the literature [36,59–61].

Interestingly, the presence of MBG particles in the microporous group had little influence on the compressive properties since pPCL and pMBG-PCL had similar stress-strain curves (**Figure 6C**) and modulus, which is consistent to a previous report using microporous scaffold

[36], demonstrating that the microporosity was the determining factor for variations in mechanical properties. The incorporation of inorganic filler generally increases the

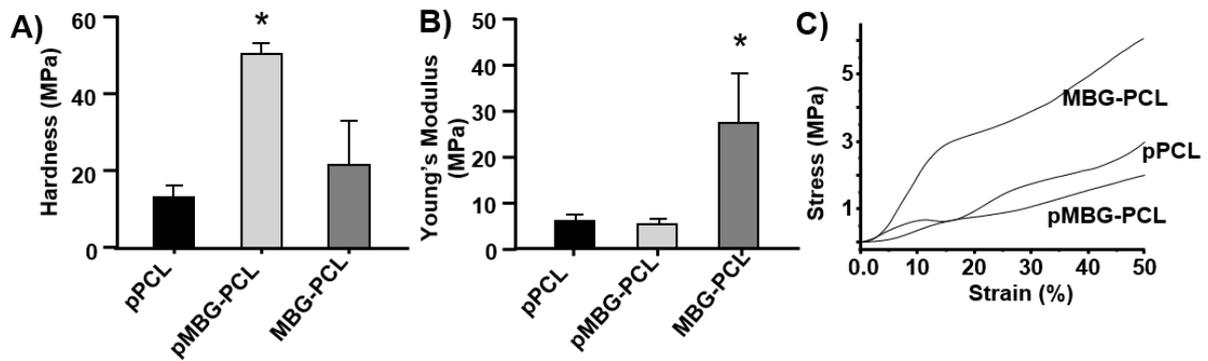

**Figure 6.** A) Hardness, B) Compressible Young's modulus and C) representative stress-strain curves of the scaffolds after the treatment in NaOH. The star represents statistical ($p < 0.05$) to all other groups (n=5).

compressive modulus of additively manufactured scaffolds [61], however this effect was negligible when microporosity was present which may have changed the deformation behaviour of the scaffold. Indeed, the micropores most likely collapsed upon the application of compression in accordance to Gibson and Ashby theory of foam compressive deformation before the MBG could experience any significant loading [62]. While the biomechanical properties were drastically reduced when compared to non-microporous scaffolds, the pPCL and pMBG-PCL remained relatively stiff despite exhibiting a 30% intra-strut porosity. This is in contrast to the softer sponge-like behaviour of a similar PCL microporous scaffold previously reported, which is likely to be related to a higher porogen content [26]. The design of the scaffold consisted in the deposition of two consecutive layers (0/0/90/90 in order to increase the pore size of the construct while maintaining equivalent mechanical properties as previously reported for PCL scaffolds [63]. Here again, the introduction of micro-porosity in the FDM strut was the determining factor in the modification of the mechanical behaviour as our double layer scaffold displayed similar compressive modulus (around 5 MPa) than single layered microporous constructs as previously reported [36].

The in vitro bioactivity of the three different scaffolds was investigated by soaking the specimens in Simulated Body Fluid for 3 days. All groups demonstrated the formation of a biomineralised layer on their surface, albeit of highly distinct morphologies (Figure 7). The pPCL scaffolds displayed sporadically dispersed calcium phosphate nodules with small dimensions, typical of early stages of biomimetic mineralisation [64,65]. The presence of the MBG in the MBG-PCL and pMBG-PCL scaffolds favoured the formation of a biomineralised layer, although the MBG-PCL group displayed a coating made of cauliflower-like CaP onto which larger mineralised nodules and needles were attached. In contrast, the pMBG-PCL scaffolds exhibited cauliflower-like nodules homogenously distributed over the entire surface of the scaffold, which resurfaced the previous mineralised layer formed during the leaching out process. This new layer presented all the topographical characteristics of an apatite-like phase [57,66], and due to the experimental conditions of the assay (pH, SBF concentration and temperature) could be related to HA, as showed previously [67]. In addition, Figure 7 shows the FTIR spectra of the scaffold's surfaces after 1h, 24 h and 3 days. The spectra confirms the formation of CaP crystalline (band P-O doublet at 600-700 cm-1) during the immersion in SBF for pMBG-PCL and MBG-PCL. The presence of CaP was detected after 1 h of immersion for the pMBG-PCL specimens, while an incipient band is detected after 3 d of immersion in SBF for MBG-PCL specimens consistently to previous reports [33,68]. This crystalline CaP precipitation in the pMBG-PCL was probably favoured by the initial mineralization during the PBS leaching out step.

Cell metabolic activity was studied in order to evaluate the influence of the incorporation of MBG particles and the formation of the micro-porosity on cell behaviour. **Figure 8A** shows the Alamar Blue percentage of reduction of M3CT3-E1 after 4, 7 and 14 days. The metabolic activity is significantly higher for the sample pMBG-PCL, probably due to both the increased surface area created by the microporosity and the mineralisation layer, which has been shown to enhance cell activity and proliferation [69,70].

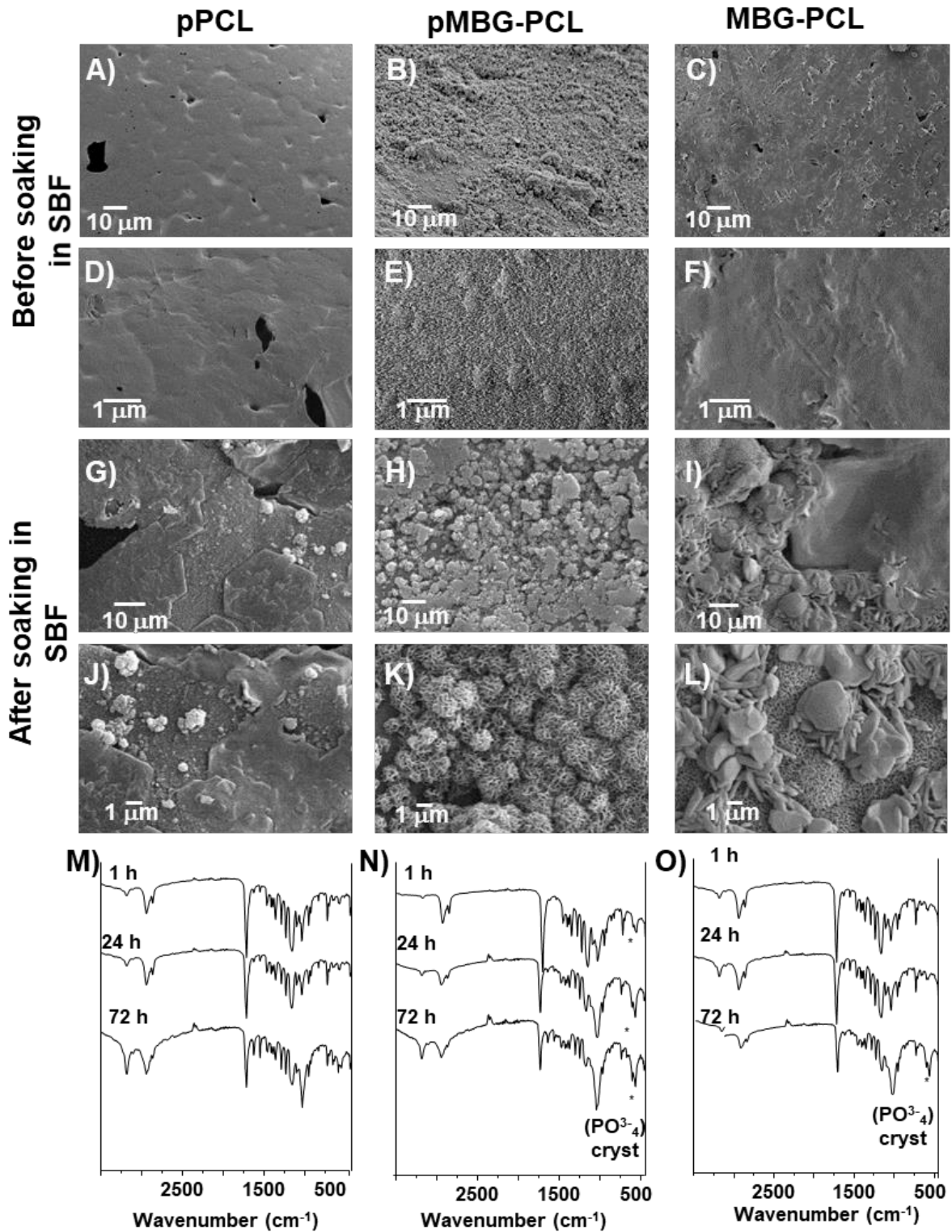

**Figure 7.** SEM micrographs A)-F) pPCL, pMBG-PCL and MBG-PCL scaffolds surface before soaking in SBF and G)-L) after 3 days of immersion in SBF. FTIR spectra M) pPCL, N) pMBG-PCL and O) MBG-PCL after 1 h, 24 h and 72 h of immersion in SBF.

**Figure 8B** shows the intracellular ALP activity as a marker of osteogenic differentiation of M3CT3-E1 pre-osteoblasts. After 7 days of incubation, the ALP activity was significantly higher in MBG-PCL when compared to pPCL and pMBG-PCL. There was no significant difference in the ALP secretion between these last two groups. At 14 days post seeding, ALP production was significantly increased in all groups compared to the previous timepoint. Interestingly, ALP secretion was significantly higher in the MBG-PCL and pMBG-PCL when compared to pPCL at this later timepoint. The MBG containing groups had similar ALP production, indicating that microporosity did not play a significant role in osteogenic commitment of the pre-osteoblasts.

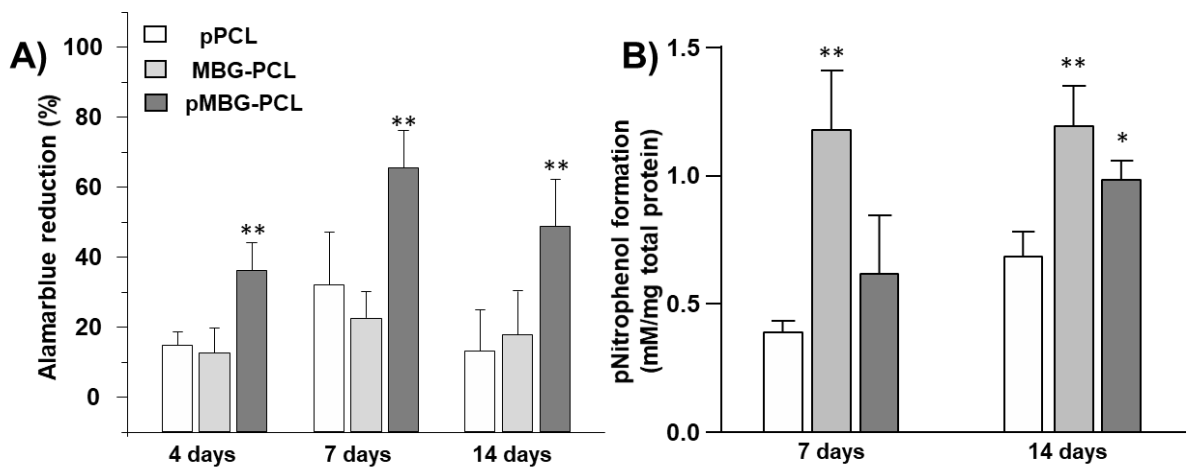

**Figure 8.** A) Cell Metabolic activity of MC3T3-E1 cells at 4, 7 and 14 days in basal medium, B) ALP activity at 7 and 14 days in osteogenic medium (**<0.01 and *<0.05 show statistical significance compared to the normalized pPCL at the corresponding timepoint) (n=5).

**Figure 9** shows confocal and SEM images of the scaffolds after 7 and 14 days of culture. The SEM images revealed that the cells had spread evenly on the pMBG-PCL scaffold surface and displayed a morphology typical for cells of an osteoblastic lineage. At 14 days post-seeding, the cells had colonised most of the scaffold and a continuous cell layer was formed in all of the groups, thus demonstrating continuous cell proliferation and confirmed the Alamar Blue findings. Confocal microscopy images exhibited similar results, that is, the complete coverage of the scaffold by cells at 7 days for pMBG-PCL scaffolds and in all of the groups after 14 days of culture.

The level of osteogenic commitment of the cells was also assessed by immuno-fluorescence targeting RUNX2, as shown in **Figure 10 A**. MCT3T3-E1 were cultured for 14 days in basal and osteogenic media to study the differences in the expression of the osteoblast differentiation associated transcription factor Runt-related transcription factor-2 (RUNX2). As expected, the results showed a significant increase in the intracellular RUNX2 concentration in response to osteogenic media when compared to basal medium for all groups. Interestingly, the microporosity combined with the presence of MBG enhanced RUNX2 expression in basal the medium when compared to both pPCL and non-microporous MBG-PCL. In addition, the presence of MBG particles alone also significantly increased RUNX2 expression, as the MBG-PCL group was higher than the pPCL in basal medium. In osteogenic medium, RUNX2 expression was significantly increased in the presence of MBG for both micro-porous and non microporous groups, indicating that the bioceramic was the determining factor for osteogenic commitment in osteogenic medium, which is consistent with previous reports [44].

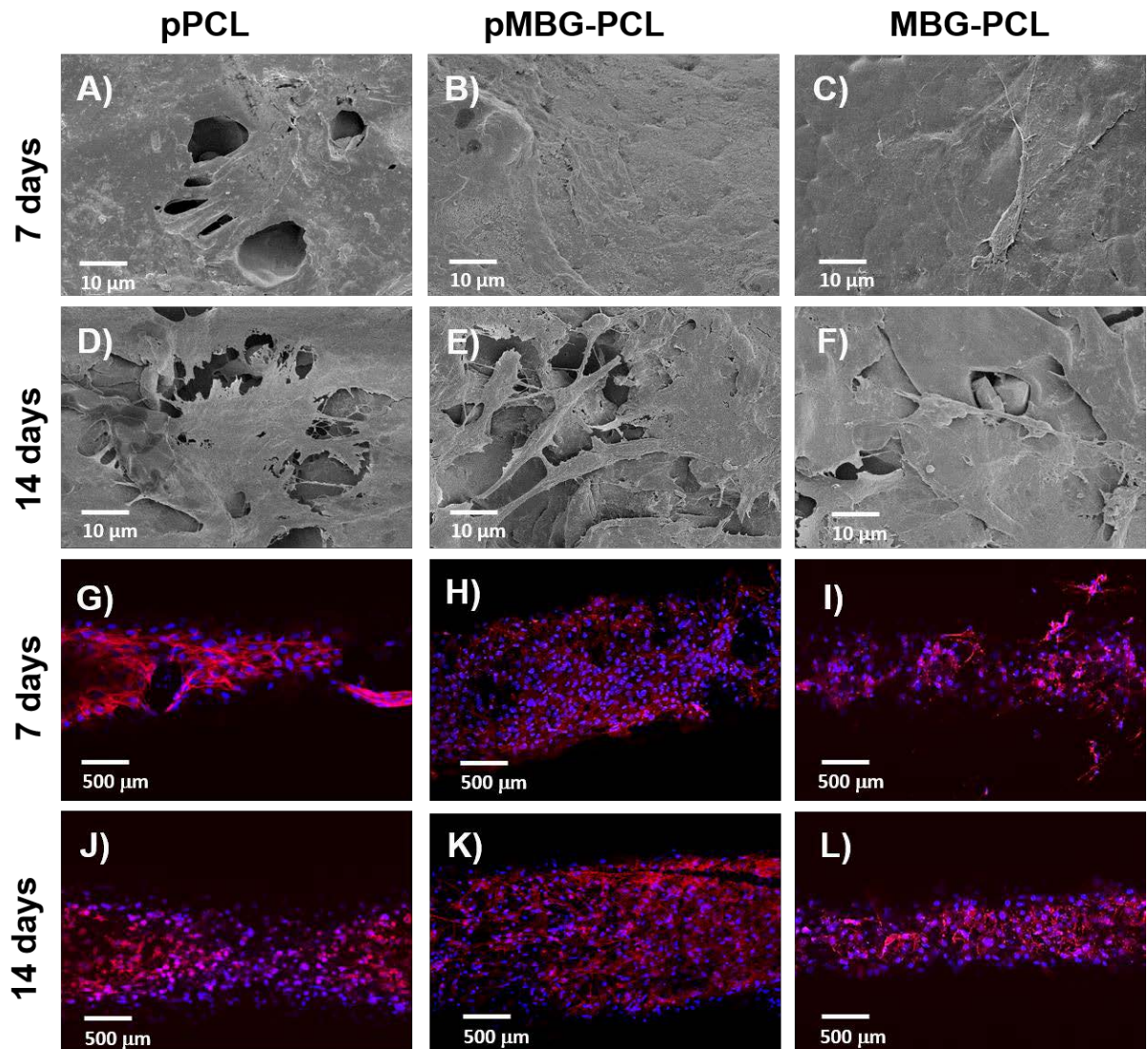

**Figure 9.** SEM micrographs of MC3T3E-1 cells on the different surfaces A)-F) and confocal images of the scaffolds G)-L) after 7 and 14 days of cell culture in basal media (n=2).

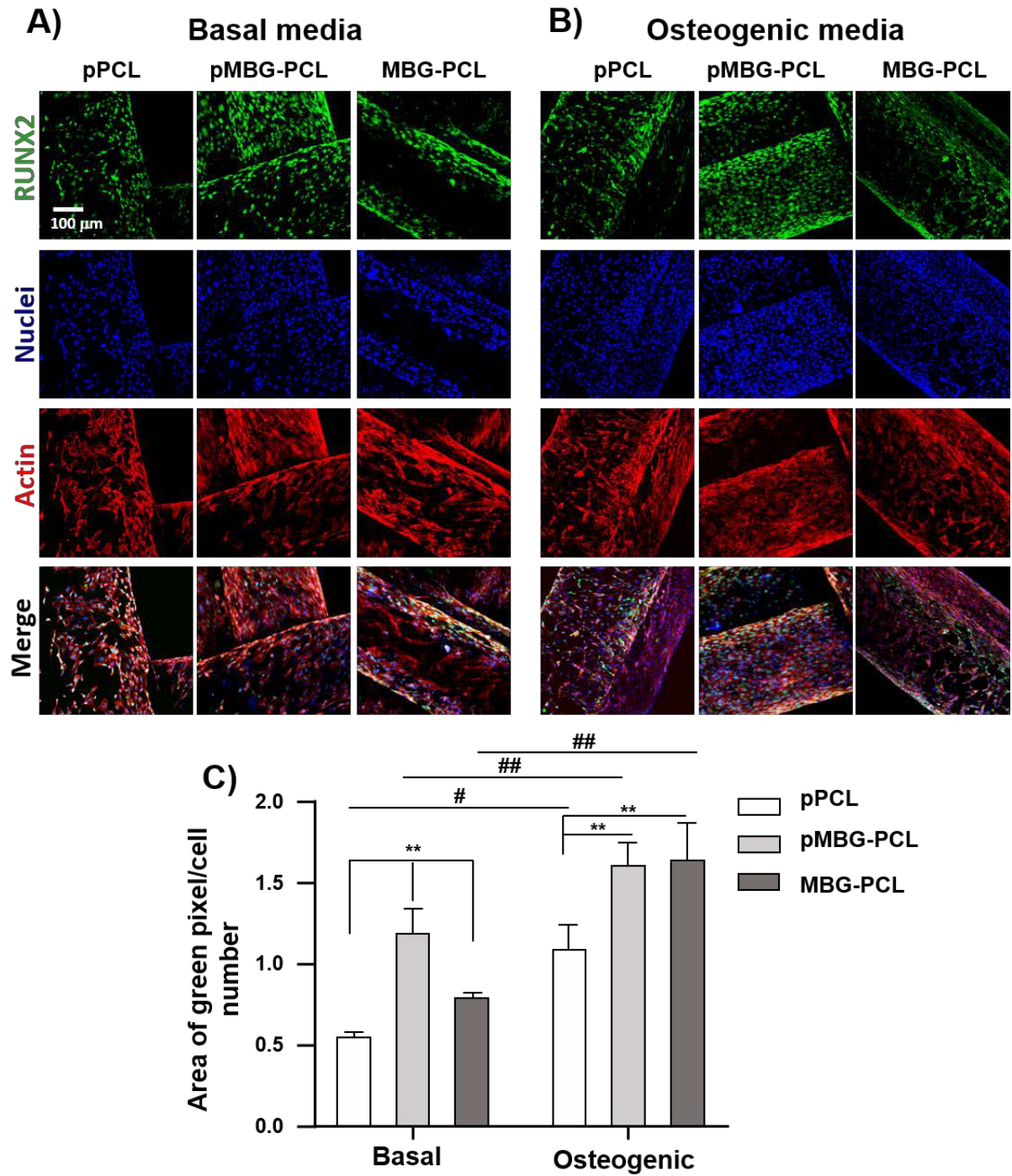

**Figure 10.** RUNX2 immunostaining images (green), nuclei (blue) and actin (red) of MC3T3-E1 on 3D printed scaffolds A) in basal media and B) in osteogenic media. C) Relative quantification of RUNX2 after 14 days of culture (n=3). (**<0.01 and *<0.05 show statistical significance between the groups at the same media and ##<0.01 and #<0.05 show statistical significance between samples cultured in different media.

## 4. Conclusions

This study reported on the development of a multiscale porous scaffold whereby macro-, micro- and meso-porosity were created using 3D-printing, porogen leaching and the incorporation of mesoporous bioglass particles. The resultant scaffold displayed a highly interconnected macroporosity inherent to the additive manufacturing technique used, which is suitable for facilitating neo-vascularisation and bone formation. The microporosity was obtained by the removal of 40 micron porogens, which created a microporous network of 30% porosity while the mesoporosity was imparted by the presence of the MBG particles. The concomitant effects of the leaching of the PBS and the release of calcium from the MBG particles resulted in the formation of a crystalline calcium phosphate layer over the pMBG-PCL and significantly increased surface hardness. It was demonstrated that the modification in the biomechanical properties was driven by the introduction of the microporosity, as opposed to the incorporation of inorganic fillers that is conventionally seen in non-microporous 3D-printed scaffold. The additional porosity increased cell metabolic activity due to the increase in surface area available for cellular interaction of the microporous scaffold and was further enhanced by the presence of the MBG particles.

## Acknowledgments

M.V.R. acknowledges funding from the European Research Council (Advanced Grant VERDI; ERC-2015-AdG Proposal No. 694160). The authors also thank to Spanish MINECO (MAT2016-75611-R AEI/FEDER, UE). The authors acknowledge the facilities, and the scientific and technical assistance, of the Australian Microscopy & Microanalysis Research Facility at the Centre for Microscopy and Microanalysis, The University of Queensland.